\documentclass{moriond}
\usepackage{amsmath,amssymb}
\usepackage{appendix}
\usepackage{comment}
\usepackage{bbold}
\usepackage{xcolor}
\usepackage{color}
\usepackage{slashed}
\usepackage{subfigure}
\usepackage{setspace}
\usepackage{footnote}
\usepackage{multirow}
\usepackage{braket}
\usepackage[normalem]{ulem}
\usepackage[utf8]{inputenc}
\usepackage{mathrsfs}
\usepackage{longtable}
\usepackage{enumitem}

\def\eV{{\rm eV}}

\def\MeV{{\rm MeV}}
\def\kpc{{\rm kpc}}

\def\be{\begin{equation}}
\def\ee{\end{equation}}
\def\bea{\begin{eqnarray}}
\def\eea{\end{eqnarray}}

\newcommand{\Photo}{}

\begin{document}
\vspace*{4cm}
\title{Constraining pseudo-Dirac neutrinos from a galactic core-collapse supernova}

\author{ Manibrata Sen }

\address{Max-Planck-Institut f{\"u}r Kernphysik, Saupfercheckweg 1, 69117 Heidelberg, Germany}

\maketitle\abstracts{
Neutrinos can be pseudo-Dirac in nature --  Majorana fermions behaving  as Dirac fermions for all practical purposes. In such a scenario, active and sterile neutrinos are quasi-degenerate in mass, and hence oscillations between the two, due to their tiny mass-squared difference $(\delta m^2)$, can develop only over very long baselines.  Under this hypothesis, we analyze the neutrino data from SN1987A, and find a mild preference for a non-zero mass-squared difference. The same data can also be used to exclude values of $\delta m^2\sim 10^{-20}{\rm eV}^2$ -- the smallest constrained so far.  We also discuss how next-generation experiments like the DUNE and Hyper-Kamiokande  can probe this scenario for a future galactic supernova.}
%%%%%%%%%%%%%%%
\section{Introduction}
%%%%%%%%%%%%%%%
The origin of neutrino masses has been eluding the particle physics community for decades. Broadly, neutrino masses can either be categorized into Dirac, or Majorana, according whether lepton number is a conserved symmetry of the Standard Model (SM) or not. In fact, probing lepton number violation is one of the very few ways to distinguish between these two hypothesis, since in the relativistic limit, Dirac and Majorana neutrinos behave identically in oscillation experiments. However, it is also possible that while lepton number is violated in the SM, it happens in a controlled, soft manner, such that for all practical purposes, neutrinos behave as Dirac fermions, although they are originally Majorana fermions.  These pseudo-Dirac (PD) fermions are characterized by a maximal mixture of almost degenerate active and sterile states. Such a tiny mass-squared difference $\delta m^2$ allows for the possibility of oscillations between active-sterile neutrinos over very long baselines.

Since these active-sterile oscillations take place over a long distance $\propto \left(E_\nu/\delta m^2\right)$, the strongest bounds come from astrophyical sources like solar neutrinos~\cite{deGouvea:2009fp}($\delta m^2\lesssim 10^{-12}\eV^2$), atmospheric neutrinos~\cite{Beacom:2003eu}($\delta m^2 \lesssim 10^{-4}\eV^2$), and even smaller values ($\delta m^2\sim 10^{-24}\eV^2$) can be probed by a measurement of the diffuse supernova background neutrinos~\cite{DeGouvea:2020ang}. Of particular interest to this scenario is a galactic core-collapse supernova (SN), which emits almost all of its energy in the form of ${\cal O}(\MeV)$ neutrinos. The relatively small energy of the neutrinos, coupled with the baseline of around  ${\cal O}(\kpc)$, makes it a natural laboratory to test oscillations driven by tiny $\delta m^2$.

In this proceeding based on \cite{Martinez-Soler:2021unz}, we will discuss how one can use neutrino observation data from SN1987A by the Kamiokande-II (KII) collaboration~\cite{Kamiokande-II:1987idp}, the IMB collaboration~\cite{Bionta:1987qt} and the Baksan collaboration~\cite{Alekseev:1988gp} to probe active-sterile oscillations arising in the PD hypothesis. Using an unbinned likelihood analysis, we find that the combined data from all the three experiments have a slight preference for active-sterile oscillations, over the no-oscillation hypothesis. On the other hand, one can use the same observations to constrain a mass-squared difference,  $ \delta m^2 \sim 10^{-20}\eV^2$, the smallest value probed so far. Finally, we also discuss how future experiments like the Hyper-Kamiokande (HK) and the Deep Underground Neutrino Experiment (DUNE) can utilize a galactic SN to constrain even tinier values of $ \delta m^2$, thereby providing stringent bounds on the PD hypothesis.

%%%%%%%%%%%%%%%
\section{Pseudo-Dirac Neutrinos}\label{sec:PDN}
%%%%%%%%%%%%%%%
The SM can be minimally extended by adding three right handed neutrinos $(N)$. After electroweak symmetry breaking, the neutrino mass matrix $(M_\nu)$ consists of a Dirac mass term, given by  $Y v/\sqrt{2}$ (where $v/\sqrt{2}$ denotes the Higgs vacuum expectation value, and $Y$ refers to the Yukawa matrix), and a Majorana mass $(M_N)$ term. The limit of soft lepton-number violation is achieved by setting $M_N\ll Y v$, such that $M_N$ breaks the degeneracy between the masses of the left and right handed states. In such a limit, $M_\nu$ can be diagonalized by a $6\times 6$ unitary  matrix, consisting of two diagonal unitary matrices: the PMNS matrix $U$, and another unitary matrix $U_N$ that diagonalizes the sterile sector~\cite{Kobayashi:2000md}. A neutrino flavor field $\nu_{\beta L}$ ($\beta =e,\mu,\tau$) can be written as a superposition of an active and a sterile field, such that $\nu_{\beta L}=U_{\beta k}(\nu_{ks}+i\,\nu_{ka})/\sqrt{2}$ for $k=1,2,3$. Note that due to the smallness of $M_N$, the states are maximally mixed, and have quasi-degenerate masses $m_{ks,ka}^2 = m_k^2 \pm \delta m_k^2/2$.  Henceforth, for simplicity, we will drop the subscript $k$, and assume that the mass difference is the same for all states, given by $\delta m^2$.

If the $\delta m^2$ is much smaller than the mass-squared differences associated with the solar and atmospheric differences, oscillations due to the latter are averaged out over astrophysical distances, whereas those due to $\delta m^2$ survive. The oscillation probability between $\nu_\beta$ and $\nu_\gamma$, driven by $\delta m^2$, can be computed as,
\begin{align}\label{eq:Prob}
    P_{\beta\gamma} = \frac{1}{2}\left(1+e^{-\left(\frac{L}{L_{\rm coh}}\right)^2}\cos\left(\frac{2\pi L}{L_{\rm osc}}\right)\right)\times\sum_{k} \left|U_{\beta k}\right|^2 \left|U_{\gamma k}\right|^2\,,
\end{align}
where $ L_{\rm osc} = (4\pi E_\nu)/(\delta m^2)$ indicates the oscillation length for a neutrino of energy $E_\nu$. Neutrinos traversing astrophysical baselines can be sensitive to decoherence due to separation of wavepackets after a certain distance, given by $ L_{\rm coh} = (4\sqrt{2} E_\nu^2 \sigma_x) /(|\delta m^2| )$, where  $\sigma_x$ measures the size of the neutrino wave-packet.
%%%%%%%%%%%%%%%
\section{Application to neutrino data from SN1987A}
%%%%%%%%%%
The neutrino fluence from a SN at a distance $d$ can be approximated by the alpha-fit spectra~\cite{Tamborra:2012ac},
\begin{align}\label{eq:Flux}
    \phi_\beta(E_\nu) = \frac{{\cal E}_{\rm tot}}{E_{0\beta}}\frac{1}{4\pi d^2}\frac{(1+\alpha)^{1+\alpha}}{\Gamma(1+\alpha)}\left(\frac{E_\nu}{E_{0\beta}}\right)^\alpha e^{-(1+\alpha)\frac{E_\nu}{E_{0\beta}}},
\end{align}
where $E_{0\beta}$ is the average energy for $\nu_\beta$, and ${\cal E}_{\rm tot}$ gives the total energy emitted in ergs. Here, $\alpha=2.3$ determines the width of the distributions, and we fix $d=50\,$kpc. 

To test our hypothesis, we consider that the fluence in Eq.\,\ref{eq:Flux} gets processed by the PD probability in Eq.\,\ref{eq:Prob}, and fit it to the $\mathcal{O}(30)$ events observed in KII, IMB and Baksan in total. We perform an unbinned extended likelihood $(\mathscr{L})$ analysis for each experiment, by carefully adopting the background treatment \cite{Vissani:2014doa}. We take effectively two flavors $e$, and $x$, which can be a linear combination of $\mu$ and $\tau$. We fit the parameters $\{{\cal E}_{\rm tot}, E_{0e}, E_{0x}, \delta m^2\}$, while $\sigma_x=10^{-13}$~m is kept fixed~\cite{Kersten:2015kio}.

%%%%%%%%%%%%%%%%%%%%%%%%%%%%%%%%%%%%%%%%%
\begin{figure*}[t!]
\centering
\includegraphics[width=0.5\textwidth]{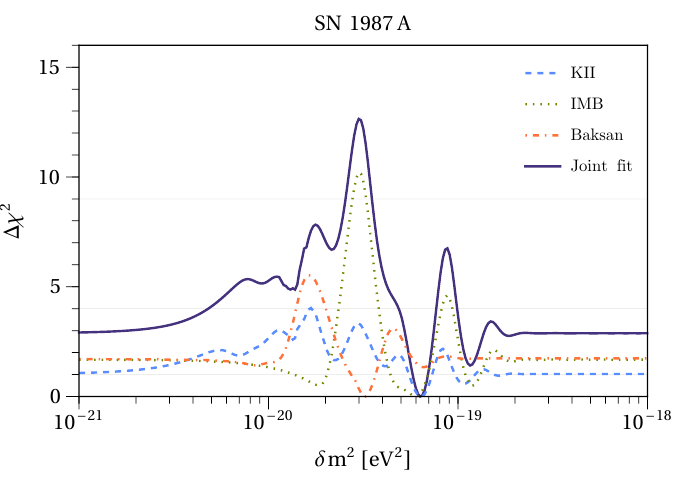}~~\includegraphics[width=0.5\textwidth, height=0.35\textwidth]{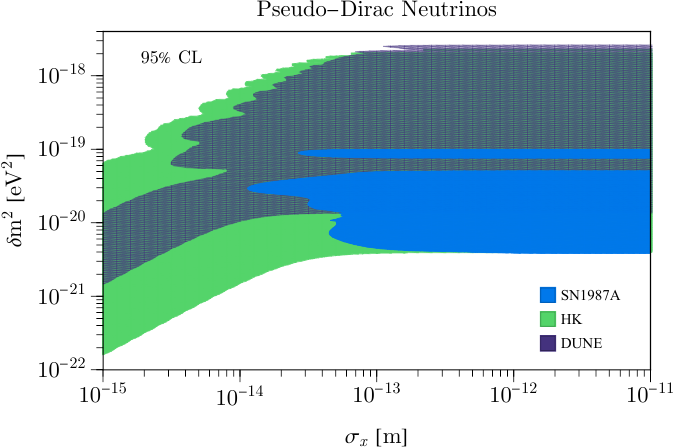}
\caption{Left: Plot showing marginalized $\Delta\chi^2$ as a function of $\delta m^2$ for KII, IMB, Baksan, and the combined set of three experiments. Right: Sensitivity plot on the $\delta m^2-\sigma_{x}$ plane for DUNE and HK, along with the exclusion contours from SN1987A for comparison.\label{fig:chi287}}
\end{figure*}
%%%%%%%%%%%%%%%%%%%%%%%%%%%%%%%%%%%%%%%%%

In the left panel of Fig.~\ref{fig:chi287}, we demonstrate the results of our fit through $\Delta \chi^2\equiv -2(\ln{\mathscr{L}} - \ln{\mathscr{L}_{\rm max}})$, which is calculated as a function of $\delta m^2$, by marginalizing over the other parameters. In this analysis, values of $\delta m^2<10^{-21}\eV^2$ correspond to no active-sterile oscillations, as the oscillation lengths are longer than what is relevant for a galactic SN. We find that KII (light-blue dashed) prefers a non-zero $\delta m^2$ at a $\Delta\chi^2\approx 1.1$ over the non-oscillation hypothesis. This is primarily because KII observed zero events in the energy range between $( 21\MeV , 31\MeV)$, which is better explained in the PD hypothesis due to the presence of an oscillation minima. Similarly IMB (green dotted) and  Baksan (orange dot-dashed) also prefer the PD scenario by a $\Delta\chi^2\approx 1.7$, although there is some tension between their best fit values. 

Finally, the combined data from all three experiments prefer a non-zero value of $\delta m^2 = 6.31\times 10^{-20}\eV^2$, disfavoring the no-oscillation scenario with $\Delta\chi^2\approx 3$. This is primarily dominated by  the tension in data between KII and IMB. Active-sterile oscillations due to $\delta m^2$ allow to reconcile this difference, by predicting a broader spectra for IMB, while leading to a suppression of events observed in KII. Furthermore, the data can be used to exclude values  $2.55\times 10^{-20}\eV^2\lesssim \delta m^2 \lesssim 3.0\times 10^{-20}\eV^2$ with a $\Delta\chi^2 > 9$. This makes it the lowest values of $\delta m^2$ constrained by experiments till date, and provides the strongest constraints on the PD hypothesis. However, one needs to do a more dedicated Monte-Carlo analysis, before assigning a statistical significance to the exclusion region.
%%%%%%%%%%%
%%%%%%%%%%%%%%
\subsection{Sensitivity studies for a future galactic supernova}
%%%%%%%%%%%%
%%%%%%%%%%%
This brings us to an important question -- in the event of a future galactic SN, where we are far better prepared to detect thousands of events, how will the bounds on $\delta m^2$ improve?  For the analysis, we consider DUNE and HK, because of their large fiducial volume, and excellent energy reconstruction. We assume the SN to take place at $10\kpc$. For the analysis, we treat ${\cal E}_{\rm tot}$ and $E_{0}$ as free parameters, while $\delta m^2$ is varied for a series of values of $\sigma_x$. This allows us to get the sensitivity to both oscillations over large distances, as well as to decoherence due to wave-packet separation.

Our results are depicted in the right panel of Fig.~\ref{fig:chi287}, which shows the $95\%$ sensitivity contours for DUNE (purple) and HK (green).  We find that, for the baseline chosen, the maximum sensitivity appears around $\delta m^2 \sim 10^{-20}\eV^2$. For smaller values of $\delta m^2$, oscillation baselines are longer than $10\kpc$, and hence sensitivity decreases. However, for $\sigma_{x} \leq 10^{-14}\text{m}$, some sensitivity still exists since a lower value of $\sigma_{x}$ cancels the effect of low $\delta m^2$. On the other hand, for a larger $\delta m^2$, decoherence nullifies the effect of rapid oscillations. Furthermore, for values of $\sigma_{x}>10^{-13}\text{m}$, there is little sensitivity to values of $\sigma_{x}$ since the coherence length remains larger than the relevant baseline. We find that HK and DUNE results can be complementary, since HK can probe smaller values of $\delta m^2$ due to its massive size, whereas DUNE has more sensitivity to larger values of $\delta m^2$ due to its good energy resolution. To compare with existing results, the region which can be explored using data from SN1987A is shown in blue. Both DUNE and HK can easily probe a large chunk of the excluded region from SN1987A, and therefore, will be able to strengthen the results significantly.

%%%%%%%%%%%%%%%%%
\section{Conclusion}
%%%%%%%%%%%%%%%%%
The signatures of pseudo-Dirac neutrinos can be manifested through active-sterile oscillations over long baselines. In this respect, we explored the scenario that neutrinos are pseudo-Dirac, and studied the impact on the neutrino fluence from SN1987A. Curiously, we found that the existing data from SN1987A presents a mild preference for the pseudo-Dirac hypothesis. Furthermore, using the same data, we excluded the tiniest values of mass-squared differences considered so far, $2.55\times 10^{-20}\eV^2\lesssim \delta m^2 \lesssim 3.0\times 10^{-20}\eV^2$, with a $\Delta\chi^2 > 9$. Finally, we used next generation experiments like DUNE and HK to study the constraints that can be imposed on this scenario from a future galactic core-collapse supernova. Needless to say, the occurrence of a galactic supernova will open up a plethora of avenues for exploration in neutrino physics.

%%%%%%%%%%%%%%%%
\section*{Acknowledgments}
%%%%%%%%%%%%%%%
We thank Ivan Martinez-Soler and Yuber Perez-Gonzalez, in collaboration with whom this work was performed. We would like to thank the Instituto de Fisica Teorica (IFT UAM-CSIC) in Madrid for support via the Centro de Excelencia Severo Ochoa Program under Grant CEX2020-001007-S, during the Extended Workshop “Neutrino Theories”, where this manuscript was completed.

\end{document}